\documentclass[aps,prl,twocolumn,floatfix, amsmath, amssymb]{revtex4-2}
\usepackage[utf8]{inputenc}
\usepackage{times}
\usepackage{physics}
\usepackage{graphicx}
 \usepackage{color}
 \usepackage{soul}
 \usepackage{bm}
 \usepackage[colorlinks=true,linkcolor=blue]{hyperref}

\newcommand{\bX}{{\bm X }}

\newcommand{\bphi}{{\bm \phi}}

\newcommand{\bc}{{\bm c}}

\newcommand{\vphi}{\varphi}

\newcommand{\bfS}{{\bf S}}
\newcommand{\bfX}{{\bf X}}

\newcommand{\hp}[1]{\textcolor{black}{{#1}}}
\newcommand{\rev}[1]{\textcolor{black}{{#1}}}
\newcommand{\rrev}[1]{\textcolor{black}{{#1}}}

\begin{document}
\title{Bloch-Landau-Zener oscillations in a quasi-periodic potential} 

\author{Henrique C. Prates}
 \affiliation{
 	Departamento de F\'isica and Centro de F\'isica Te\'orica e Computacional, Faculdade de Ci\^encias, Universidade de Lisboa, Campo Grande, Edif\'icio C8, Lisboa 1749-016, Portugal 
}

\author{Vladimir V. Konotop}
 \affiliation{
 	Departamento de F\'isica and Centro de F\'isica Te\'orica e Computacional, Faculdade de Ci\^encias, Universidade de Lisboa, Campo Grande, Edif\'icio C8, Lisboa 1749-016, Portugal 
}
\begin{abstract}
    \rev{Bloch oscillations and Landau-Zener tunneling are ubiquitous phenomena which are sustained by a band-gap spectrum of a periodic Hamiltonian and can be observed in dynamics of a quantum particle or a wavepacket in a periodic potential under action of a linear force. Such physical setting remains meaningful for aperiodic potentials too, although band-gap structure does not exist anymore. Here} we consider the dynamics of noninteracting atoms and Bose-Einstein condensates in a quasi-periodic one-dimensional optical lattice subjected to a weak linear force. Excited states with energies below the mobility edge, and thus localized in space, are considered. We show that the observed oscillatory behavior is enabled by tunneling between the initial state and a state (or several states) located nearby in the coordinate-energy space. The states involved in such Bloch-Landau-Zener oscillations are determined by the selection rule consisting of the condition of their spatial proximity and condition of quasi-resonances occurring at avoided crossings of the energy levels.  The latter condition is formulated mathematically using the Gershgorin circle theorem. The effect of the inter-atomic interactions on the dynamics can also be predicted on the bases of the developed theory. The reported results can be observed in any physical system allowing for observation of the Bloch oscillations, upon introducing incommensurablity in the governing Hamiltonian.
\end{abstract}

\maketitle
Bloch oscillations and Landau-Zener tunneling are fundamental phenomena predicted nearly a century ago~\cite{Bloch,Zener,LandauTun}. Nowadays, they are experimentally observed for electrons in solids~\cite{FeLeSh,Leo1992,Waschke1993}, cold atoms~\cite{Peik1997a,Peik1997b, BhMaMo} and Bose-Einstein condensates (BECs) in optical lattices (OL)~\cite{Morsch2001,Cristiani2002,Ferrari2006,Gustavsson2008,Kling2010,Geiger2018, AndKas, BrWiKo}, arrays of optical waveguides~\cite{Peschel1998,Pertsch1999,Morandotti1999, Dreisow2009}, periodic dielectric media~\cite{Sapienza2003}, optical resonators~\cite{Yuan2016}, exciton-polaritons in microcavities~\cite{Flayac2011}, plasmonics~\cite{Block2014}, \rev{acoustic fields in water cavities~\cite{Sanchis2007} and phononic crystals~\cite{Lanzillotti2010}, 
and in non-Hermitian systems~\cite{Longhi2009,Xu2016}}  (see also  reviews~\cite{Nenciu1991,Kolovsky2004}). In virtue of the ubiquity of these phenomena and possibilities of modifying experimental settings toward including aperiodic potentials, further attention was paid to effect of randomness on Bloch oscillations~\cite{DAMaMoLy03, ScDrKl, DrKlWi, Stutzer2013}, as well as to dynamics of particles in deterministic aperiodic media~\cite{MoLyDAMa05, MoLyDAMa08, WaSchDu, Wang14, ReGaBe, YaSaSh, Kang2008}.  
Despite previous studies, the known results for tilted quasi-periodic potentials, remain scarce, limited to extended states, and somewhat controversial. 

Indeed, for a particle in a deep tilted lattice, where aperiodicity is treated as a perturbation and description relies on usage of the conventional Bloch-band theory, numerical studies of~\cite{MoLyDAMa05} revealed sustained dynamics while in~\cite{WaSchDu} the oscillations were found damped. In Ref.~\cite{Wang14} investigating Harper's tilted lattice, where Bloch theory is not applicable anymore, fragmentation of oscillations in momentum and in real spaces, as well as sensitivity of the dynamics to the initial state and to the system parameters, were reported. On the other hand, even though the mobility edge (ME)~\cite{Mott} is known to exist for certain types of quasi-periodic potentials~\cite{Sarnak1982,Suslov1982,Ostlund1983,FroSpWi} and was mentioned in~\cite{MoLyDAMa05,Wang14} in the context of Bloch oscillations, only extended states were addressed, so far. Among the available results we also mention experimental observation of the oscillatory dynamics of a BEC in a weakly disordered quasi-periodic lattice~\cite{ReGaBe}, and numerical studies of Bloch oscillations in two-dimensional quasi-periodic lattices~\cite{MoLyDAMa08,YaSaSh}.

In this Letter, we report on the dynamics of atoms and BECs in a one-dimensional tilted quasi-periodic OL in the regime when tight-binding approximation is not applicable. We concentrate on the states having energies below the ME and thus, localized in the coordinate space. {In the absence of Bloch-band spectrum, the motion of the atoms, referred below as Bloch-Landau-Zener (BLZ) oscillations, is sustained by simultaneous tunneling in the coordinate space and inter-level transitions enabled by the tilt [see Fig.~\ref{fig1} (a)]. The tunneling events can be viewed as quasi-resonances where the resonant modes are chosen according to the  {\em selection rule}. This rule stems from intersection of Gershgorin circles~\cite{Gershgorin} and from the condition of non-negligible spatial hopping of the wavefunctions.

To describe BLZ oscillations we consider the dimensionless Gross-Pitaevskii equation for the wavefunction $\Psi$ (normalized to one):
\begin{equation}\label{eq:SE}
    i\partial_t \Psi
    =H_\alpha \Psi+g \abs{\Psi}^2 \Psi. 
\end{equation}
Here $H_\alpha=H_0-\alpha x$, $H_0=-(1/2)\partial_x^2+V(x)$ is the linear Hamiltonian in the absence of tilt, and $\alpha$,  $0<\alpha\ll 1$, is the strength of the linear force. For definiteness below we concentrate on a quasi-periodic 
potential $V(x)=v\cos{(2x)}+ v\cos{(2\varphi x+\theta)}$, where $v$ is the amplitude and $\varphi$ is an irrational relation between the periods of the sub-lattices. An arbitrarily chosen $\theta$ breaks spatial symmetry, and $g$ characterizes the strength of inter-atomic interactions. Bi-chromatic incommensurate potentials with sub-lattices of comparable amplitudes were studied in Refs.~\cite{Diener01, PraZezKon22,YaKhBrSP, BoGoHiHo, Azbel79}. When one of the sub-lattices has much larger amplitude, the tight-binding approximation becomes applicable and the model is reduced to an Aubry-Andr\'e-Harper type equation~\cite{AubryAndre, Kohmoto83, Modugno09, BiWaPrSar, Molina2014}. This limit will not be considered here. For cold atoms  such potential was produced experimentally in~\cite{Luschen2018}.   

The spectrum of $H_0$ with the quasi-periodic potential $V(x)$ on the whole real axis is of particular complexity~\cite{Simon82,FroSpWi,Surace1990}. Meantime a real-world condensate has a finite spatial dimension. Therefore the model can be simplified by considering a BEC in an infinite potential well~(see e.g.\cite{Gaunt2013,Navon2021}) whose width $L$ greatly exceeds the periods of the sublattices: $L\gg 1$. This means that $x\in [-L/2,L/2]$  and zero boundary conditions must be  imposed: $\Psi (\pm L/2,t)=0$ (for localized states centered far from the boundaries, the assumption that $L$ is finite is not important~\cite{supplementary}). Now our goal can be re-formulated: given an initial ($t=0$) localized eigenstate of $H_0$ we explore its dynamics at $t>0$ for $\alpha>0$.   

To this end we consider the eigenvalue problem $H_\alpha\phi_n^{(\alpha)}(x)=\epsilon_n^{(\alpha)}\phi_n^{(\alpha)}(x)$, where $n$ numbers the  orthonormal eigenstates (they are nondegenerate):  $\langle\phi_n^{(\alpha)},\phi_m^{(\alpha)}\rangle=\delta_{mn}$ (thereafter $\langle f,g\rangle=	\int_{-L/2}^{L/2}f^*(x)g(x)dx$), and $n=1$ corresponds to the ground state. 
Starting with the linear case, $g=0$,  for $L$ large enough, we can define~\cite{Diener01,Modugno09,PraZezKon22,YaKhBrSP,BoGoHiHo, ZezKon22} a ME $\epsilon_{\rm ME}$ such that all eigenstates of $H_0$ with $\epsilon_n^{(0)}<\epsilon_{\rm ME}$ ($\epsilon_n^{(0)}>\epsilon_{\rm ME}$)  are localized (delocalized) [see Fig.~\ref{fig1} (b)]. Here the localization  means that the characteristic width of the state, measured as $\chi_n^{-1}$, where $\chi_n=\langle |\phi_n^{(0)}|^2,|\phi_n^{(0)}|^2\rangle$ is the inverse participation ratio (IPR), is much less than the size of the system:  $1/\chi_n\ll L$. 
\begin{figure}[htb]
 \centering
\includegraphics[width=\linewidth]{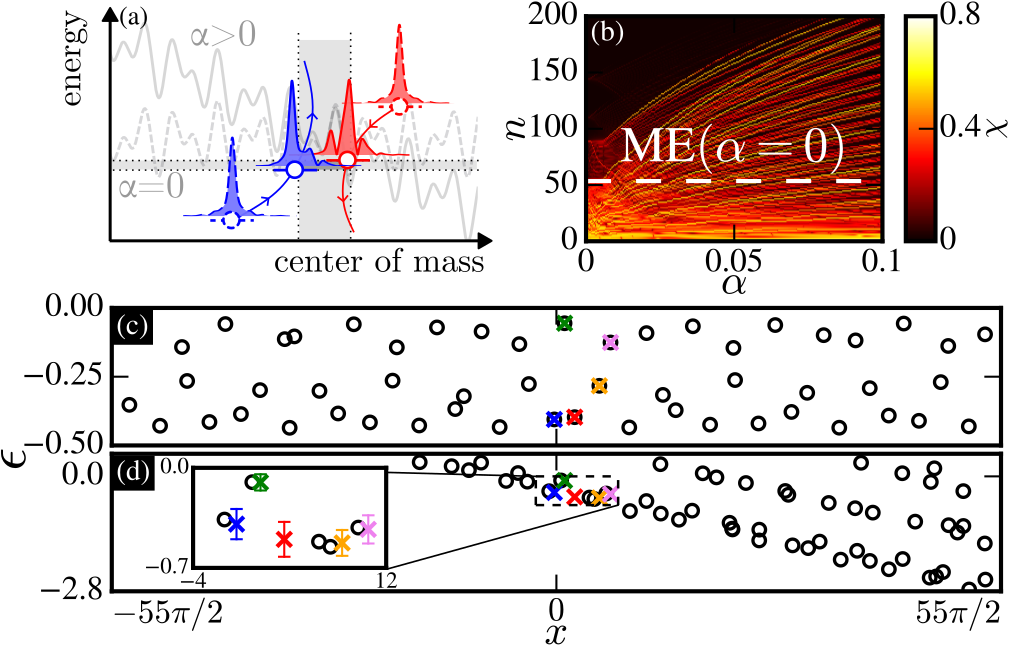}
\caption{(a) Schematics of the selection rule. Two states created at $\alpha=0$ (dashed lines) when $\alpha$ increases move in the coordinate-energy space along trajectories shown by arrows. These states do not interact until for certain $\alpha_1$ they approach each other (solid lines) closer than a distance ensuring non-negligible hopping and proximity of the energies (intersection of shadowed stripes). At $\alpha\approx \alpha_1$ the states are in quasi-resonance.  
Grey lines illustrate the potential without (dashed) and with (solid) tilt. (b) IPRs of the eigenstates of $H_\alpha$. The dashed line indicates the ME for $\alpha=0$.  Energies and centers of masses (empty circles) of states with $\epsilon_n^{(\alpha)}<\epsilon_{\rm ME}^{(\alpha)}$ for $\alpha=0$ and $\epsilon_{\rm ME}^{(0)}=-0.056$ (c), and $\alpha=0.029$ and $\epsilon_{\rm ME}^{(\alpha)}=0.78$ (d). The colored crosses and diamonds in (c) and (d) represent $\varepsilon_n^{(\alpha)}$ of the states involved in the dynamics shown in Fig.~\ref{fig2}. Vertical bars in the inset are the Gershgorin intervals. The parameters are $v=1$, $\vphi=(1+\sqrt{5})/2$, $\theta=1.23$, and $L=55\pi$.}
\label{fig1}
\end{figure}

Let $N$ be the total number of localized states at $\alpha=0$, i.e., $\epsilon_N^{(0)}< \epsilon_{\rm ME}< \epsilon_{N+1}^{(0)}$. A low-energy wavepacket in such lattice without tilt is given by $\Psi=\sum_{n=1}^{N}c_n^{(0)}(t)\phi_n^{(0)}(x)$. Since at $\alpha>0$ each eigenstate undergoes displacement in the coordinate-energy space [see Fig.~\ref{fig1} (a)], we consider also the "displaced" basis:  $\bphi_{\alpha}=(\phi_1^{(\alpha)}, \dots, \phi_N^{(\alpha)})^T$ 
($T$ stands for the transpose).  Strictly speaking, at $\alpha>0$ the ME smears out [see Fig.~\ref{fig1} (b)] and a possibility of atomic transfer to  extended states may occur. However, for the low-energy sates the probability of such events remains negligibly small because of smallness of the hoping integrals between localized and delocalized states. This conclusion is supported by numerical simulations that we have performed, thus justifying the use of the $N$-mode approximation for $\alpha\ll 1$. 
 
The bases $\bphi_{0}$ and  $\bphi_{\alpha}$ are connected via the unitary transformation $ \bphi_\alpha=\bfS_\alpha \bphi_0$, where the entries of the matrix $\bfS_\alpha$  are $S_{mn}^{(\alpha)}=\langle \phi_n^{(0)},\phi_m^{(\alpha)}\rangle$. Assuming that initially a state $\phi_n^{(0)}$ is prepared, our task can be reformulated as determining the states $\phi_m^{(0)}$ (with $m\neq n$) to which the atoms are transferred in a tilted lattice, and describing the respective evolution. 
  
We characterize the spatial position of a localized state with its center of mass (c.m.) which can be expressed using a basis obtained for any $\alpha$: $X(t)=\bc_{\alpha}^\dagger(t){\bf X}_{\alpha}\bc_{\alpha}(t)$, where $\bc_{\alpha}=(c_1^{(\alpha)}, \dots, c_N^{(\alpha)})^T$, and ${\bf X}_{\alpha}$ is an $N\times N$ matrix with entries $X_{mn}^{(\alpha)}=\langle \phi_m^{(\alpha)},x\phi_n^{(\alpha)}\rangle$. The diagonal element  $X_{nn}^{(\alpha)}$ can be interpreted as the c.m. of a state $\phi_n^{(\alpha)}$.  
  Different eigenstates in quasi-periodic potentials are localized at different lattice sites~\cite{ZezKon22,PraZezKon22}. In Fig.~\ref{fig1} this is illustrated for the unperturbed $H_0$ [panel (c)] and tilted $H_\alpha$ [panel (d)] Hamiltonians. 
  At $\alpha=0$ one observes a nearly homogeneous distribution over the $x$ axis, while panel (d) for $\alpha>0$ can be viewed as a Wannier-Stark ladder (for the lower states) where unlike in the cases of periodic potentials (see e.g.~~\cite{VoBlBo,WiBhMa,HeHiSt,Gluck2002}) {\em states with different energies are localized at different spatial steps.}
 
Turning to the linear dynamics, from Eq.~\eqref{eq:SE} with $g=0$ we obtain:
$id \bc_0/dt=({\bf E}_0-\alpha {\bf X}_0)\bc_0$   
and  
$	id \bc_{ \alpha}/dt={\bf E}_{ \alpha} \bc_{ \alpha},$ where ${\bf E}_\alpha=\mbox{diag}(\epsilon_1^{(\alpha)},...,\epsilon_N^{(\alpha)})$, in the unperturbed and tilted bases, respectively.
The evolution of $c_n^{(\alpha)}$ is trivial: $\bc_\alpha(t)={\bm \Lambda}(t)\bc_\alpha(0)$ where 
 ${\bm \Lambda}(t)=\mbox{diag} \left(e^{-i\epsilon_1^{(\alpha)}t},...,e^{-i\epsilon_N^{(\alpha)}t}\right)$ and in terms of the initial populations, the c.m. has the form~\cite{supplementary}
 \begin{align}\label{X0}
 	X(t)=\bc_0^\dagger(0) \bfS_\alpha^\dag{\bm \Lambda}^\dag(t)\bfS_\alpha   {\bf X}_0 \bfS_\alpha^\dag{\bm \Lambda}(t)\bfS_\alpha \bc_0(0).
\end{align}
Certain entries in this formula give negligible contributions to the dynamics. Indeed, suppose that at $t=0$ only a state $m$ is excited, i.e., $c_n^{(0)}(0)=\delta_{n m}$. Then, $c_n^{(\alpha)}(0)=S_{nm}^{(\alpha)}$, i.e., the amplitudes of the state $\phi_n^{(\alpha)}$ is determined by the hopping integral $S_{nm}^{(\alpha)}$. Most of these integrals are negligible, except those involving states satisfying a {\em spatial proximity} condition 
\begin{align}
\label{cond-ref}
    |\langle \phi_n^{(0)},\phi_m^{(\alpha)}\rangle| > \Delta_x
\end{align}
where $\Delta_x\ll 1$ is determined by the desired accuracy. 
Only such states are excited in the oscillatory motion. 

Furthermore, it follows from (\ref{X0}) that time dependence of the $X(t)$ is determined by $ \epsilon_n^{(\alpha)}-\epsilon_m^{(\alpha)}$. To express this quantity through the eigenvalues  at $\alpha=0$, we consider the diagonal elements of the matrix ${\bf E}_0-\alpha {\bfX}_0$:  $\varepsilon_n^{(\alpha)}=\epsilon_n^{(0)}-\alpha X_{nn}^{(0)}$ [for a few  states involved in dynamics described below, they are marked by colored crosses in Figs.~\ref{fig1} (c) and (d)].
Since $\epsilon_n^{(\alpha)}$ is an eigenvalue of the matrix  ${\bf E}_0-\alpha {\bfX}_0$, the Gershgorin circle theorem~\cite{Gershgorin} assures that for each $\epsilon_{n}^{(\alpha)}$ there exist $\varepsilon_m^{(\alpha)}$ such that
$|\epsilon_{n}^{(\alpha)}-\varepsilon_m^{(\alpha)}|\leq |R_m^{(\alpha)}|$ where  $ R_m^{(\alpha)}=\alpha\sum_{m'\neq m}^N|X_{mm'}^{(0)}| $ are the Gershgorin radii. 
The subindexes $m$ and $n$ in this formula may not coincide because unlike $\epsilon_n^{(\alpha)}$ the diagonal elements $\varepsilon_m^{(\alpha)}$ considered as functions of $\alpha$ can cross [see Fig.~\ref{fig2} (a)]. We refer to such event of crossing as {\em quasi-resonance}. Notice that this quasi-resonance is essentially different from the resonances induced by driving of the Aubry-Andr\'e model~\cite{Molina2014}, as well as from recently studied resonances governing behavior of disordered systems in the many-body localized phase~\cite{Ghosh2022,Morningstar2022}. Thus, $\epsilon_{n}^{(\alpha)}$ may be found inside a Gershgorin interval (due to reality of the spectrum in our case the circles are squeezed to intervals) centered at $\varepsilon_{m}^{(\alpha)}$ originating from the eigenvalue $\varepsilon_m^{(0)}=\epsilon_m^{(0)}$.

The states below the ME are distributed in the space nearly homogeneously [Fig.~\ref{fig1} (c)], i.e., the average distance between neighboring states $\sim L/N$, while for their maximal widths one estimates $1/\chi=1/\min_{m}\{\chi_m\}$. Since $|X_{mn}^{(0)}|\lesssim 1/\chi$~\cite{supplementary}, for $1/\chi\lesssim L/N$, valid for our case, only the nearest neighbors contribute to the sum in $R_m^{(\alpha)}$. Thus, we obtain the estimate $|R_m^{(\alpha)}|\lesssim 2\alpha /\chi$. For the modes in the inset of Fig.~\ref{fig1} (d) this yields $|R_m^{(\alpha)}|\lesssim 0.2$.

Suppose now, that energies of two eigenstates of $H_0$, say $\epsilon_{m}^{(0)}$ and $\epsilon_{m'}^{(0)}$, originate Gershgorin intervals which overlap at a certain value of $\alpha$ [see Fig.~\ref{fig1} (a)], i.e.,  
\begin{align}
	\label{cond_ener1}
|\varepsilon_{m}^{(\alpha)}-\varepsilon_{m'}^{(\alpha)}|\leq R_{m}^{(\alpha)}+R_{m'}^{(\alpha)}.
\end{align}
For such states $|\epsilon_{m}^{(\alpha)}-\epsilon_{m'}^{(\alpha)}|\leq 2(R_{m}^{(\alpha)}+R_{m'}^{(\alpha)})$. If also $R_{m,m'}^{(\alpha)}$ are small enough and spatial hopping of the wavefunctions is non-negligible, they interact resonantly and can be identified as resonant. Thus, the conditions of spatial proximity (\ref{cond-ref}) and of the overlapping of Gershgorin intervals (\ref{cond_ener1}) constitute a {\em selection rule}, which allows one, for a given initial state $m$ to determine a state $m'$ of the Hamiltonian $H_0$, which becomes excited in the tilted quasi-periodic potential. The number of resonant states {\it versus} tilt is shown in Fig.~\ref{fig2} (c).
\begin{figure}
\centering
\includegraphics[width=\linewidth]{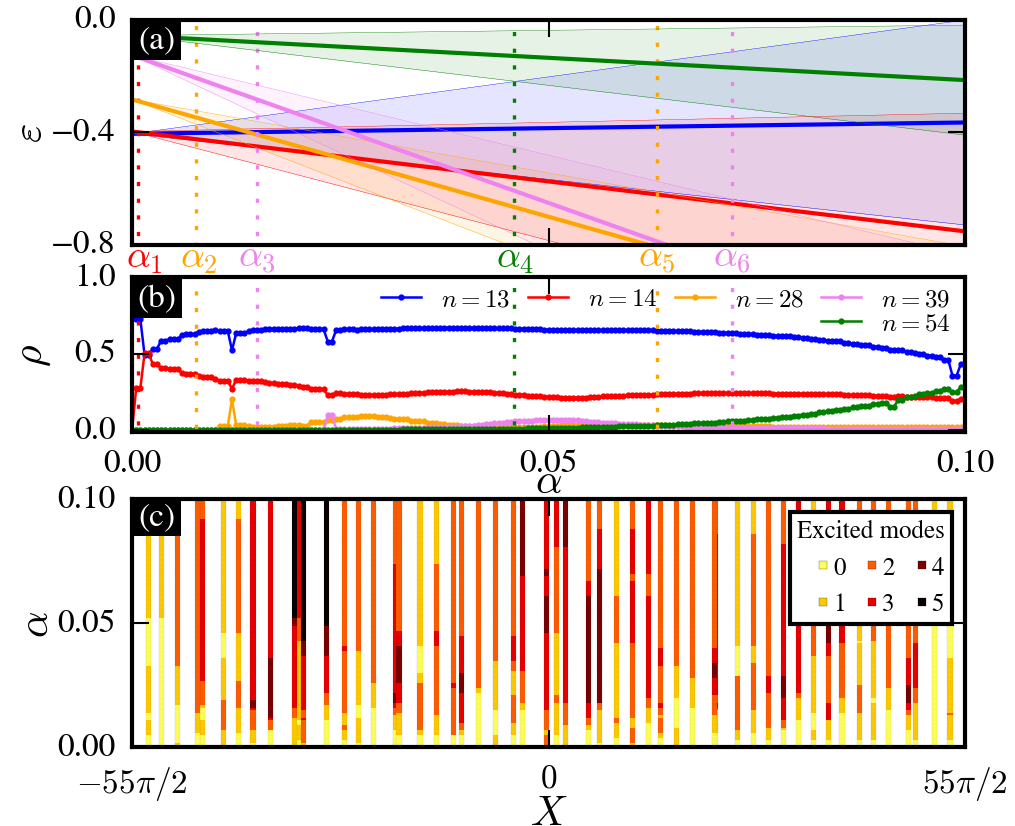}
\caption{(a) Centers $\varepsilon$ (solid lines) of Gershgorin intervals (shaded area) {\it versus} tilt. (b) Average populations of excited states obtained from ~(\ref{eq:SE}) with  $\Psi_0=\phi_{13}^{(0)}$  for the evolution time $T=10^3$.
 The vertical dotted lines indicate $\alpha$ for which the Gershgorin intervals intersect ($\alpha_1$, ..., $\alpha_4$) and stop intersecting ($\alpha_5$, $\alpha_6$).
 (c) Numbers of states satisfying the selection rule at $\alpha>0$ obtained for the states which for $\alpha=0$ are centered at $X_{nn}^{(0)}$ (corresponding to $X$ coordinate of colored stripes whose finite widthsare  used for better visibility).
The potential is the same as in Fig.~\ref{fig1}. }
\label{fig2}
\end{figure}

The selection rule for sufficiently small $\alpha$ admits a transparent physical interpretation. Indeed, for strongly localized modes the Gershgorin radii determined by the hopping integrals are small and the quasi-resonance,  $\varepsilon_{m}^{(\alpha)}=\varepsilon_{m'}^{(\alpha)}$ can be interpreted as approximate equality $\epsilon_m^{(0)}-\epsilon_{m'}^{(0)}\approx \alpha (X_{mm}^{(0)}-X_{m'm'}^{(0)})$. This means that two states $m$ and $m'$ are involved in  atomic exchange if the work of the linear force necessary for spatial transfer between their c.m. is approximately equal to the energy difference between the states.  The described dynamics   corresponds to {\em simultaneous inter-level transitions in the energy space and tunneling in the coordinate space}. 

Turning to numerical study of the evolution governed by Eq.~\eqref{eq:SE} with the initial condition $\Psi(x,t=0)=\Psi_0(x)$, we characterize participation of a state $n$ in the dynamics by its average population: ${\rho}_n=\frac{1}{T}\int_0^T |\langle \phi_n^{(0)},\Psi_0\rangle|^2 dt$ where $T$ is the total time of the evolution. Figure~\ref{fig2} (b) illustrates the results for $\Psi_0(x)=\phi_{13}^{(0)}(x)$ shown by the blue cross in Fig.~\ref{fig1} (c). In close proximity to that state one can find a localized state $\phi_{14}^{(0)}$ [red cross in Fig.~\ref{fig1} (c)]. As a result, already at $\alpha=\alpha_1\approx 0.0007 $ tunneling between these states occurs. Tunneling between these states remains for the whole interval $\alpha\in[0,0.1]$ [see intersecting blue and red Gershgoring intervals in Fig.~\ref{fig2} (a)]. Respectively, in Fig.~\ref{fig2} (b) one observes that the mentioned states (blue and red lines) remain populated for all $\alpha\in[0,0.1]$. For $\alpha<\alpha_2\approx 0.0079$ (the left yellow dotted line) there are no other excited states satisfying the selection rule, and periodic oscillations [Fig.~\ref{fig3} (a)] occur only due to tunneling between the two states.  At exact quasi-resonances corresponding to the intersection of solid lines in Fig.~\ref{fig2} (a) when $\varepsilon_{n}^{(\alpha)}=\varepsilon_{m}^{(\alpha)}$, sharp peaks in the average population of the excited states occur [panel (a)], indicating that the states exchange by the largest amount of atoms.  

Upon increase of the linear force new states may satisfy the selection rule (\ref{cond-ref}), (\ref{cond_ener1}), thus becoming excited and involved in the interchange of atoms through BLZ tunneling. 
In Fig.~\ref{fig2}, this is the state  $\phi_{28}^{(0)}$ [yellow crosses in Fig.~\ref{fig1}] whose Gershgorin interval starts to overlap with that of the state $\phi_{13}^{(0)}$ at $\alpha_2$. Respectively, for $\alpha>\alpha_2$ one observes oscillatory dynamics characterized by involvement of three states and thus by two main frequencies of oscillations [see Fig.~\ref{fig3} (b)] . Further increase of the linear force results in the involvement of even more states in the evolution. For instance, at $\alpha=\alpha_3\approx 0.016 $ the state $\phi_{39}^{(0)}$ becomes involved in the population exchange process [see dynamics in \hp{Fig.~\ref{fig3}(c)}]. Remarkably, the "inverse" process is also possible: when the discs separate upon the increase of $\alpha$ the tunneling between the respective states becomes suppressed. For example, this happens for the states $\phi_{28}^{(0)}$ and $\phi_{39}^{(0)}$ at $\alpha_5\approx0.062$ and $\alpha_6\approx0.072$ \hp{[marked in Fig.~\ref{fig2}(b) by the right yellow and pink dotted lines]}, respectively. \hp{Therefore in Fig.~\ref{fig2}(b), one observes that these states remain unpopulated for $\alpha>\alpha_5$ and $\alpha>\alpha_6$.} The respective BLZ oscillations become "simpler" as shown in Fig.~\ref{fig3} (d) where the tunneling to $\phi_{28}^{(0)}$ and $\phi_{39}^{(0)}$ is strongly suppressed.

The dynamics in a quasi-periodic potential strongly depends on the choice of the initially populated state. In Fig.~\ref{fig2} (c) we show the number of states which are excited in a tilted potential provided 
that at $\alpha=0$ a localized state with a given c.m. $X_{nn}^{(0)}$ (vertical colored lines) is created. The shown numbers correspond all excited states satisfying the selection rule~\eqref{cond-ref}, ~\eqref{cond_ener1}. Notice, that there are states remaining out of resonances (marked as 0 excited states), meaning that no oscillations of such states are observed below certain values of $\alpha$. At the same time, while increase of the tilt involves more states in the dynamics, this is not the only possible scenario: states may become non-resonant upon increase of $\alpha$, because  their Gershgorin intervals become separated.

All patterns in Fig.~\ref{fig3} are non-decaying. This is also  explained by the selection rule. Indeed, absence of overlap of Gershgoring intervals means a negligibly small tunneling rate which, loosely speaking, excludes excitation of those states at any time. For the same reason for a given $\alpha$, the dynamics is well described by a few-mode approximation, i.e., by the reduced column-vector $\bc_0$ in which only excited states (of all $N$ existing) are relevant [the number of states can be determined from Fig.~\ref{fig2} (c)]. For example,  for $\alpha<\alpha_2$ in Fig.~\ref{fig2} (a), the vector $\bc_0$ has two components and $\bX_0$ is a $2\times2$ matrix (see~\cite{supplementary}). The dynamics illustrating this case is shown in Fig.~\ref{fig3} (a). Other panels of this figure show evolution which is well described by three-mode [Fig.~\ref{fig3} (b) and (d)]  and four-mode [Fig.~\ref{fig3} (c)] models. Comparing the evolution of the c.m. obtained from the full dynamics (dashed green line) and from the respective few-mode approximations (dotted blue line) reveal excellent agreement in all these cases. Notice that while several states are involved in the dynamics, which thus must be characterized by several frequencies, only one main frequency is observed in all panels. This is explained by smallness of other frequencies~\cite{supplementary} which do not manifest themselves during the evolution time shown in the figure.
\begin{figure} 
\centering
\includegraphics[width=\linewidth]{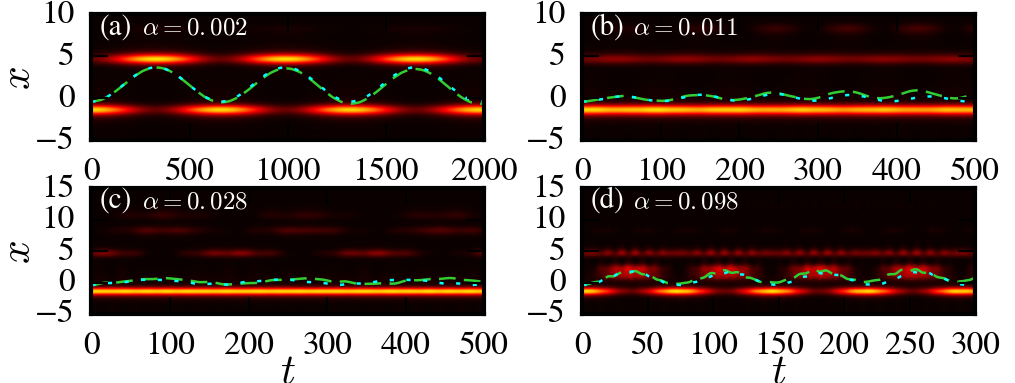}
\caption{Evolution of $\abs{\Psi}^2$ for the initial state $\Psi_0=\phi_{13}^{(0)}$ and different field strengths (indicated in the panels).  The dashed green and dotted blue line shows evolution of the c.m. obtained from Eq.~(\ref{eq:SE}) and the few-mode model. Other parameters are as in Fig.~\ref{fig1}. In (b)-(d) only few periods are shown, while simulations were carried till $t=1000$.}
\label{fig3}
\end{figure}
 
The developed theory allows one to predict the effect of weak and even moderate nonlinearity on BLZ oscillations described by (\ref{eq:SE}). A small positive scattering length $g>0$, favoring delocalization, is expected to recruit more adjacent (in the coordinate-energy space) states, thus increasing effective Gershgorin intervals. This naturally results in more complex (compared with the linear limit) dynamics, as shown in Fig.~\ref{fig4} (a) [cf. Fig.~\ref{fig3} (a)]. At larger scattering lengths relatively fast dispersion of the initial wavepacket is observed (see~\cite{supplementary} for examples). 

\begin{figure} 
\centering
\includegraphics[width=\linewidth]{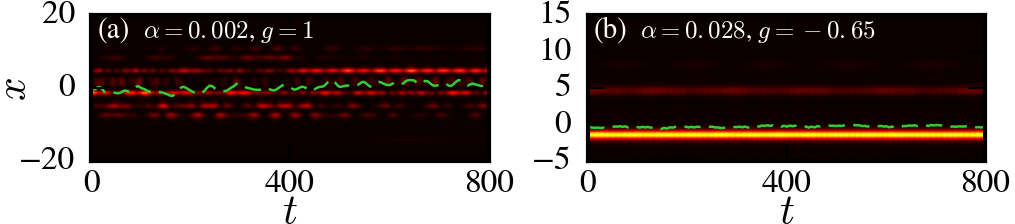}
\caption{Evolution of $\abs{\Psi}^2$ for the initial state $\Psi_0=\phi_{13}^{(0)}$ for positive (a) and negative (b) scattering lengths, corresponding to the linear dynamics in Figs.~\ref{fig3} (a) and (c).
}
\label{fig4}
\end{figure}
On the other hand, a negative scattering length ($g<0$) results in stronger localization, and thus, in smaller effective Gershgorin radii leading to suppression of the tunneling and creation of weakly oscillating matter solitons. This is illustrated in Fig.~\ref{fig4} (b) where barely visible oscillations of the c.m. are displayed [cf. Fig~\ref{fig3} (c)]. At larger intensities of the attractive two-body interactions a nearly static soliton is created (not shown here).  

 To conclude we have described sustained Bloch-Landau-Zener oscillations of atoms and BECs in a tilted quasi-periodic lattices. We considered states having energies below the mobility edge. The oscillatory motion occurs due to tunneling of atoms between an initial localized state and other localized state (or states) satisfying the selection rule. This rule consists of the  conditions of proximity of the states in coordinate and in energy spaces, achieved due to action of the applied linear force. Depending on the initial condition and on the magnitude of the tilt of the quasi-periodic potential, one can detect two- or a few- mode dynamics, or even no dynamics at all. The reported results can be observed in any of physical systems where Bloch oscillations have been realized, since quasi-periodic potentials can be obtained using essentially the same technical tools as periodic ones. 
\begin{acknowledgments}
The authors acknowledge financial support from the Portuguese Foundation for Science and Technology (FCT) under Contracts UIDB/00618/2020 (DOI: 10.54499/UIDB/00618/2020) and PTDC/FIS-OUT/3882/2020 (DOI: 10.54499/PTDC/FIS-OUT/3882/2020). HCP was supported under the FCT doctoral grant 2022.11419.BD.
\end{acknowledgments}

\end{document}


\title{Supplemental Material: Bloch-Landau-Zener oscillations in quasi-periodic potentials}

\author{Henrique C. Prates}
\affiliation{
	Departamento de F\'isica and Centro de F\'isica Te\'orica e Computacional, Faculdade de Ci\^encias, Universidade de Lisboa, Campo Grande, Edif\'icio C8, Lisboa 1749-016, Portugal 
}
\author{Vladimir V. Konotop}
\affiliation{
	Departamento de F\'isica and Centro de F\'isica Te\'orica e Computacional, Faculdade de Ci\^encias, Universidade de Lisboa, Campo Grande, Edif\'icio C8, Lisboa 1749-016, Portugal 
}
\begin{abstract}
\end{abstract}

\maketitle

\section{Effect of the choice of bounadry conditions}
\begin{figure}[h]
    \centering
    \includegraphics[width=0.99\linewidth]{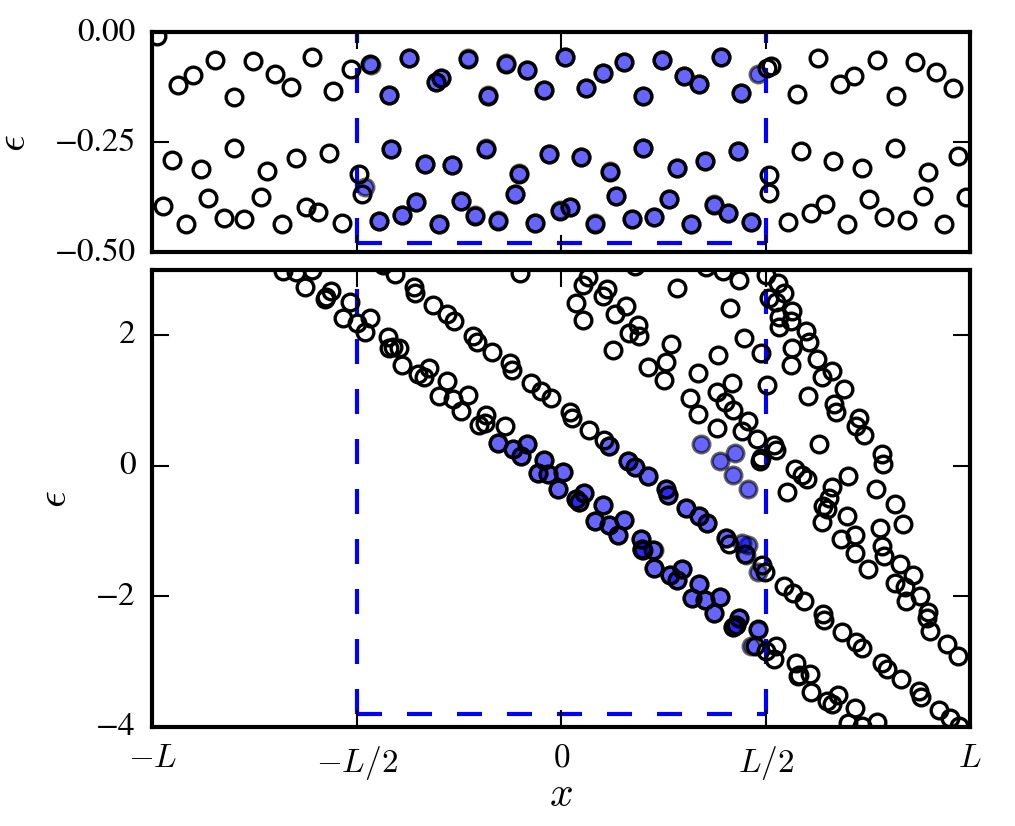}
    \caption{Distribution of the modes of in the coordinate-energy plane for $\alpha=0$ (a) and $\alpha=0.029$ (b), computed for the boxes $[-L/2,L/2]=[-55\pi/2,55\pi/2]$ (blue circles) and $[-55\pi,55\pi]$ (black circles. The dashed lines mark the potential wells of smallest box. The parameters of the potential are the same as in the main text, i.e.,  $v_1=v_2=1$, $\theta=1.23$ and $\varphi=(1+\sqrt{5})/2$.}
    \label{fig:s1}
\end{figure}

When the size of the box $[-L/1,L/2]$ increases, almost all new modes appear at spatial locations outside the previous potential well. 
In Fig.~\ref{fig:s1} we illustrated this by comparing the location of center of mass of the modes of the unperturbed Hamiltonian $H_0$, calculated for boxes of the sizes $L=55\pi$ (explored in the main text) and $L'=2L=110\pi$. On the scale of the figure one does not observe differences for the modes in the interval $x\in[-L/2,L/2]$, which are located sufficiently far from the boundaries $x=\pm L/2$.

\section{On center-of-mass displacement}

To provide some details of the derivation of the center of mass displacement  $X(t)$
used in the main text, we notice that given a basis $\bphi_{\alpha}$ and the corresponding projections $\bc_{\alpha}(t)$,  one obtains
\begin{align}
    X(t)=\braket{\Psi}{x\Psi}=&\bc_{\alpha}^\dagger(t){\bf X}_{\alpha}\bc_{\alpha}(t)
    \nonumber\\ 
    =&\bc_{\alpha}^\dagger(0) {\bf \Lambda}^\dagger (t) {\bf X}_{\alpha}{\bf \Lambda}(t)\bc_{\alpha}(0).
\end{align}
Since the bases $\bphi_{0}$ and  $\bphi_{\alpha}$ are connected via the unitary transformation $ \bphi_\alpha=\bfS_\alpha \bphi_0$, the projections are related as $\bc_\alpha(t)=\bfS_\alpha \bc_0(t)$, hence 
    ${\bf X}_{\alpha}=\bfS {\bf X}_0 \bfS^\dagger$ 
this leads to Eq.~(2) from the main text.

\section{On a few mode approximation}

When only few-modes are excited during the evolution, a few-mode approximation provides reasonably accurate description, based on the respective eigenstates of the unperturbed Hamiltonian. This reduces the Schr\"{o}dinger equation to 
\begin{equation}\label{eq:matrix_fm}
    i \frac{dc_0}{dt}=\mathbf{H} c_0, \quad \mathbf{H}_{nm}=\varepsilon_{n}\delta_{nm}-\alpha (1-\delta_{nm})X_{nm}^{(0)}.
\end{equation}
For example, the dynamics shown in  Fig.~3 (a) of the main text, when only two modes $\phi_{n_1}^{(0)}$ and $\phi_{n_2}^{(0)}$ with $n_1=13$ and $n_2=14$ is described by the Hamiltonian
\begin{align}
    \mathbf{H}=
    \left( \begin{matrix}
        \varepsilon_{n_1} & -\alpha X_{n_1 n_2}^{(0)}\\
        -\alpha X_{n_1 n_2}^{(0)} & \varepsilon_{n_2}
    \end{matrix}
    \right)
\end{align}
and the initial condition $c_{n_1}(0)=1$ and $c_{n_2}(0)=0$. It is straightforward to obtain the eigenfrequencies
 $\varepsilon_0 \pm \nu/2$ where
\begin{equation}
    \varepsilon_0=\frac{\varepsilon_{n_1}+\varepsilon_{n_2}}{2}, \quad
    \nu=\sqrt{(\varepsilon_{n_1}-\varepsilon_{n_2})^2 + 4\alpha^2 [X_{n_1 n_2}^{(0)}]^2 }.
\end{equation}
is the frequency that is observed in Fig.~3 (a); the respective period is estimated to be $2\pi/\nu\approx 659.5$.

When three states interact we have
\begin{equation}
     \mathbf{H}=
    \left( \begin{matrix}
        \varepsilon_{n_1} & -\alpha X_{n_1 n_2}^{(0)} & -\alpha X_{n_1 n_3}^{(0)}\\
        -\alpha X_{n_1 n_2}^{(0)} & \varepsilon_{n_2} & -\alpha X_{n_2 n_3}^{(0)}\\
        -\alpha X_{n_2 n_3}^{(0)} & -\alpha X_{n_2 n_3}^{(0)} &  \varepsilon_{n_3}
    \end{matrix}
    \right).
\end{equation}
In another example, shown in Fig.~3 (d) of the main text, where the interacting eigenstates are $n_1=13$, $n_2=14$ and $n_3=54$, the frequencies are computed as $\nu_1\approx-0.86$, $\nu_2\approx-0.27$ and $\nu_3\approx -0.19 $, which corresponds to two periods of LBZ oscillations $T_1\approx 73.9$, $T_2\approx 10.8$ and  $T_3\approx 9.4 $.

\section{On estimate of the Gershgorin radii}

Consider
\begin{align}
    2|X_{mn}^{(0)}|=&2\left|\int_{-L/2}^{L/2} \bar{\phi}_m^{(0)}x\phi_n^{(0)}dx\right|\nonumber\\
    =&\left|\int_{-L/2}^{L/2} \bar{\phi}_m^{(0)}(x-X_{mm}^{(0)})\phi_n^{(0)}dx \right. \nonumber\\ 
       &+\left.\int_{-L/2}^{L/2} \bar{\phi}_m^{(0)}(x-X_{nn}^{(0)})\phi_n^{(0)}dx\right|
    \nonumber \\
    &\leq \langle (x-X_{mm}^{(0)})^2\rangle^{1/2}+\langle (x-X_{nn}^{(0)})^2\rangle^{1/2}
\end{align}
where 
\begin{align}
    \langle (x-X_{mm}^{(0)})^2\rangle=\int_{-L/2}^{L/2} |\phi_m^{(0)}|^2 (x-X_{mm}^{(0)})^2 dx
\end{align}
is the mean square width of the state and we used the Cauchy inequality, as well as  the normalization and orthogonality of the states. Since in a typical situation
\begin{align}
    \langle (x-X_{mm}^{(0)})^2\rangle\sim \frac{1}{\chi_m^2}
\end{align}
we estimate
\begin{align}
    |X_{mn}^{(0)}|\lesssim \max\{\chi_m^{-1},\chi_n^{-1}\}\leq \max_{m}\{\chi_m^{-1}\}.
\end{align}

\section{Further examples of dynamics}
\begin{figure}[h]
    \centering
    \includegraphics{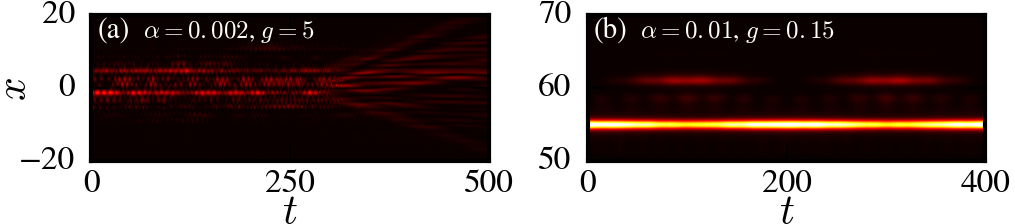}
    \caption{Nonlinear evolution of the density $\abs{\Psi}^2$ using the initial state $\Psi_0=\phi_{13}^{(0)}$ (a) and $\Psi_0=\phi_{2}^{(0)}$ (b) under the applied fields strengths $\alpha$ and nonlinearities marked in the panels.}
    \label{fig:S2}
\end{figure}
In the main text we mention large positive scattering
lengths lead to relatively fast dispersion of the initial wavepacket. We illustrate this fact in Fig.~\ref{fig:S2}, where we consider the nonlinear dynamics for the initial conditions $\Psi(x, 0)=\Psi_0(x)$. In Fig. (a) we show the nonlinear dynamics for $\Psi_0(x)=\phi_{13}^{(0)}(x)$, whose linear dynamics is illustrated in Fig. 3(a) of the main text, and undergoes complex multi-mode dynamics under sufficienctly weak positive scattering length [cf. Fig. 4(a) of the main text]. Under even stronger positive nonlinearities, full dispersion of the wavepacket is observed in Fig.~\ref{fig:S2}(a).
To further illustrate the effect of positive nonlinearities, in  Fig.~\ref{fig:S2}(b) we show the dynamics of $\Psi_0(x)=\phi_{2}^{(0)}(x)$. Notice that the observable two-mode oscillations are induced by the weak positive scattering lenght, as the initial state does not undergo BLZ oscillation in the linear case [cf. Fig. 2(c) of the main text].
\vfill

 
